\documentclass[pre,twocolumn,showpacs]{revtex4}
\usepackage{epsfig}

\newcommand{\g}[1]{{\bf {#1}}}

\begin{document}

\title{Water waves over a strongly undulating bottom}
\author{V.~P. Ruban}
\email{ruban@itp.ac.ru}
\affiliation{Landau Institute for Theoretical Physics,
2 Kosygin Street, 119334 Moscow, Russia} 

\date{\today}

\begin{abstract}
Two-dimensional free-surface potential flows of an ideal fluid over 
a strongly inhomogeneous bottom are investigated with the help of 
conformal mappings. Weakly-nonlinear and exact nonlinear equations of motion 
are derived by the variational method for arbitrary seabed shape parameterized 
by an analytical function. As applications of this theory, 
band structure of linear waves over periodic bottoms is calculated and 
evolution of a strong solitary wave running from a deep
region to a shallow region is numerically simulated.
\end{abstract}

\maketitle


\section{Introduction}

The classical problem of water waves over a variable seabed has attracted much
attention (see
\cite{Lamb,Johnson-1,LeibovichRandall,Johnson-2,Spielvogel,LozanoMeyer,
SvendsenHansen,Miles-1,Heathershaw,Mei1985,MilesSalmon,Kirby,
HaraMei,MeiHaraNaciri,
DGB1989,FIM,LiuYeh,
PutrevuOltman-Shay, LYLCC,
MilesChamberlain,Miles-2,LiuYue,Milewski,
AthanassoulisBelibassakis,Agnon,MokYeh,TAEGF2000,PMH2002,ZhenYe,PorterPorter}
and references therein). 
There are some significant differences in this interesting and
practically important problem, as compared to the theory of waves on a deep
water or in canals with a flat horizontal bottom. In situations where the
fluid depth is less or of the same order as a typical length of
surface wave,  inhomogeneity of the bottom is a reason for 
linear and nonlinear wave scattering 
and transformation, and it strongly affects wave propagation.
These phenomena occur so widely that one can meet them almost everywhere, 
although with different scales. Examples of strongly nonlinear dynamics
are ocean waves running on a beach, or motion of disturbed water 
in a puddle after a car. 
Among linear effects due to bottom topography is  existence 
of special edge-localized waves discovered by Stokes 
\cite{LiuYeh,PutrevuOltman-Shay,LYLCC,MokYeh}, 
that propagate along the shore line of a beach. Over an axially symmetric
underwater hill, quasi-localized wave modes with non-zero angular momentum
can exist, similar to long-life-time states of a quantum particle confined
by a potential barrier of a finite width \cite{LozanoMeyer,LL3}. 
It is necessary to say that underwater
obstacles of definite shapes and sizes can serve as waveguides (a narrow and
long underwater crest) or as lenses (an oblong underwater hill oriented
crosswise to the wave propagation). A qualitative explanation for all the linear
effects is simple. Indeed, let $\g{r}_\perp$ be the coordinate in the 
horizontal plane, $H(\g{r}_\perp)$ the depth corresponding to quiet surface. 
Then, looking at the well known dispersion relation for small-amplitude 
gravitational surface waves,
\begin{equation}\label{omega_K_H}
\omega(K,H)=\sqrt{gK\tanh(KH)}
\end{equation} 
(where $\omega$ is the frequency, $K$  is the absolute value of the wave 
vector, $g$ is the gravitational acceleration), one can see that the local
refraction index $n(\omega,\g{r}_\perp)$ increases as the depth 
$H(\g{r}_\perp)$ decreases, in accordance with the formulas
\begin{equation}\label{n_omega_H}
n(\omega,H(\g{r}_\perp))\equiv
\frac{K(\omega,H(\g{r}_\perp))}{K(\omega,H=\infty)}
=\frac{g K(\omega,H(\g{r}_\perp))}{\omega^2}>1,
\end{equation}
$$
\frac{\partial K(\omega,H)}{\partial H}<0,
$$
where the function $K(\omega,H)$ is determined by Eq. (\ref{omega_K_H}).
Therefore, as in the conventional light optics, 
here an oblique wave changes its
direction of propagation when meets gradient of $n$. 
Also, the total internal
reflection is possible in propagation from smaller depth to larger depth. 

Besides observing such natural phenomena, a set of laboratory experiments has
been carried out to investigate various aspects of the given problem in more
idealized and controlled conditions than are achieved in nature
\cite{MokYeh,Heathershaw,HaraMei,MeiHaraNaciri,Kirby,LYLCC,TAEGF2000}. 
In particular, waves  over locally 
periodic bottoms were studied experimentally \cite{Heathershaw,
HaraMei,MeiHaraNaciri,Kirby,TAEGF2000}, 
and such a general for periodic media effect was observed as
the Bragg resonances and the corresponding band structure with gaps in 
wave spectrum. It is worth to say that in natural conditions 
quasi-periodic sand bars occur quite often.

In general, a qualitative picture of the mentioned phenomena is clear. As
concerning the quantitative side of the mathematical theory of waves over
a variable bottom, here not everything that necessary has been done, because
practically all developed up to now analytical models and methods are 
related to the limit cases where  the fluid is considered as ideal,
and the slope of the bottom is small (or amplitude of the bottom 
undulations is small).
For the general three-dimensional (3D) Hamiltonian theory of water waves, such
restriction seems to be unavoidable even in considering the most simple,
irrotational flows when the state of the system is described by a minimal 
set of functions, namely by a pair of canonically conjugated quantities as
the deviation $\eta(\g{r}_\perp,t)$ of the free surface from the horizontal 
plane and the boundary value $\psi(\g{r}_\perp,t)$ of the velocity potential
\cite{Z68,ZK97}.
A technical difficulty exists here that, when working in 3D space, 
it is impossible to represent in convenient and compact form the kinetic energy
functional ${\cal K}\{\eta,\psi\}$ which is part of the Hamiltonian of the system. Small values of the
bottom slope and of the free surface slope make possible expansion of the
Hamiltonian to asymptotic series and subsequent application of various variants
of the perturbation theory. In such traditional approach, an inhomogeneous 
bottom does not allow to write in exact form even linearized equations, not
speaking about nonlinear corrections.

There are more favorable conditions for progress in theory of 2D
potential ideal flows with a free boundary, and the reason for this is the
possibility to employ such powerful mathematical tools as analytical functions
and the corresponding conformal mappings. Time-dependent conformal mappings were
successfully used for studying strongly nonlinear 2D wave dynamics 
on deep water and over straight horizontal bottom
\cite{Ovs,DKSZ96,DLZ95,ZD96,DZK96,D2001,ZDV2002,DZ2004}. 
In the cited works the region occupied by resting fluid (the lower half-plane 
or a horizontal stripe) was mapped onto the region with disturbed free 
boundary, and the real axis was transformed into moving boundary.
Such a conformal ``straightening'' of free surface has provided a compact 
representation for the Hamiltonian, derivation of exact equations of motion, 
and possibility for precise numerical simulations of the system evolution.

\begin{figure}
\begin{center}
  \epsfig{file=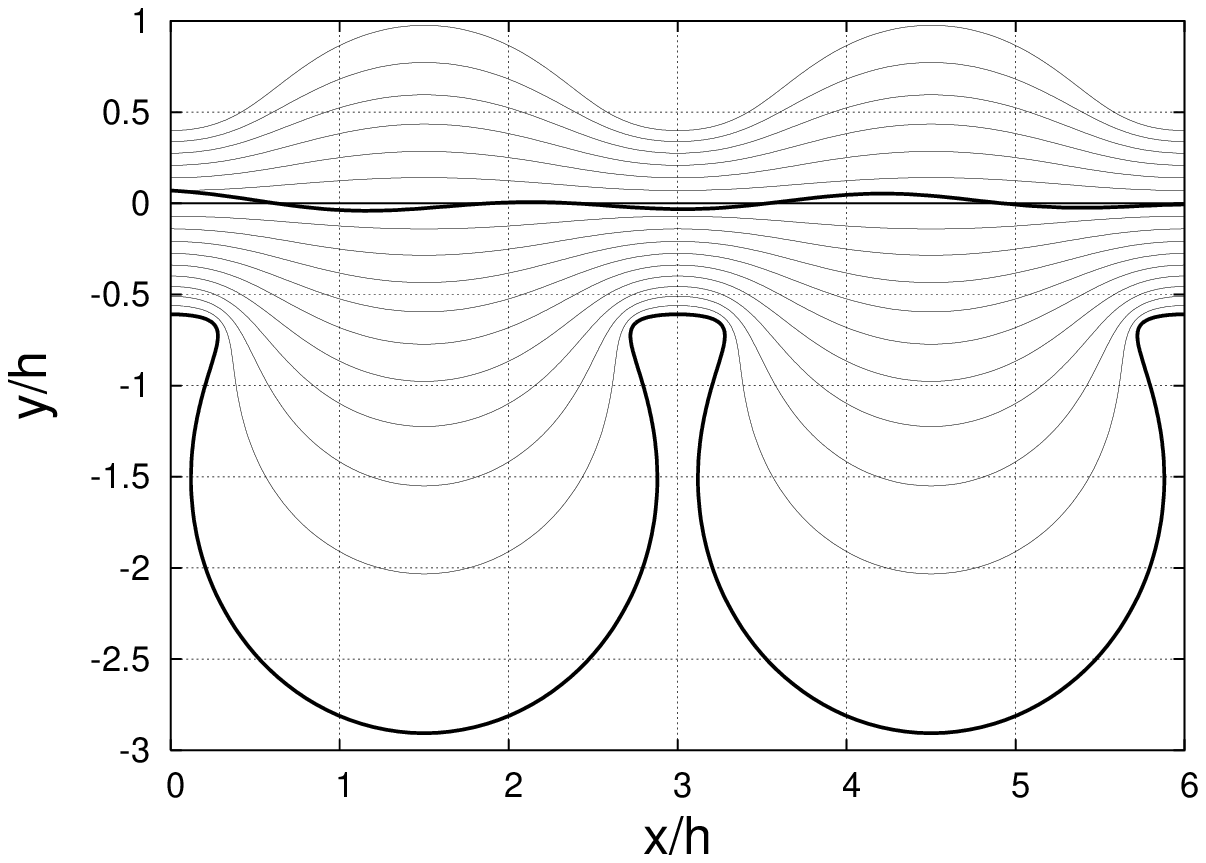,width=72mm}
  \epsfig{file=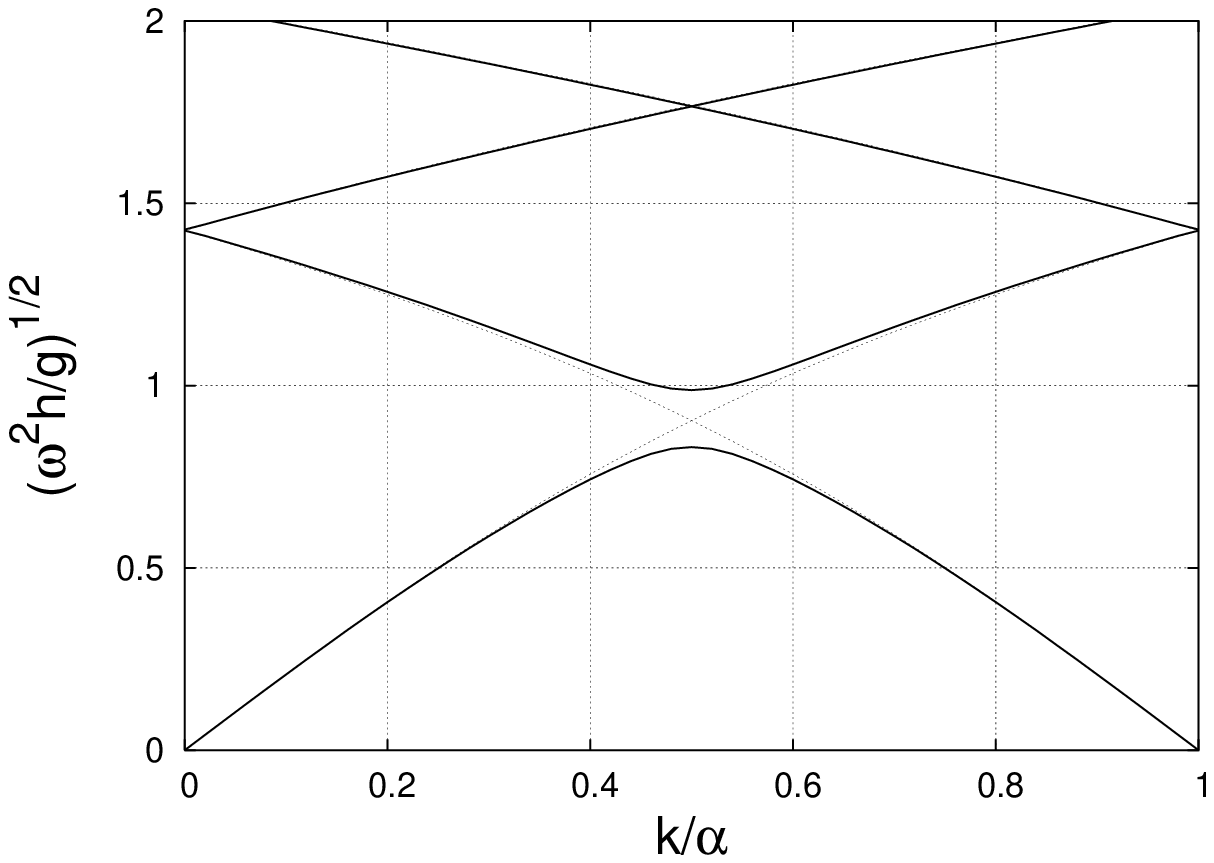,width=72mm}
\end{center}
\caption{\small Left: periodic shape of the bottom (lower thick line), 
levels of constant $v=-0.9, -0.8,...,+0.6$ (thin lines), 
and (schematically) free boundary (thick line near $y=0$). 
Right: the corresponding band 
structure of the spectrum of linear waves. In this example 
$z(w)/h=w+(2\epsilon/\alpha)\sin(\alpha w)/(1+b\cos(\alpha w))$, with the parameters
$\alpha=2\pi/3$, $\epsilon=-0.17$, $b=0.16$. } 
\label{BandStructure-1}
\end{figure}

The purpose of this work is to study the effect of a strongly undulating bottom 
on 2D ideal potential flows with a free surface. Here conformal mappings are 
used as well, and this is done in two variants. In the first, ``moderate'' 
variant (Sec. 2), a fixed conformal mapping ``straightens'' the bottom, 
but not the free boundary. More exactly: instead of the Cartesian coordinates
$x$ and $y$ (with $y$-axis up-directed), curvilinear coordinates $u$ and $v$
are introduced, and the change of coordinates is performed with the help of
an analytical function $z(w)$ which maps the stripe $-1<\mbox{Im\,} w<0$ onto the
region between the horizontal line $y=0$ and the inhomogeneous bottom 
$y=-H(x)$. In this case $x+iy=z(u+iv)$, the horizontal line $y=0$ corresponds 
to $v=0$, and on the bottom $v=-1$. The bottom may have arbitrary large slope
and even impending pieces where the dependence $H(x)$ is multi-valued, as
shown in Fig.\ref{BandStructure-1}.
The shape of free surface will be described by a function $v=V(u,t)$.
The Lagrangian for weakly-nonlinear waves is represented as an integral series
in powers of the dynamical variables $V(u,t)$ and $\psi(u,t)$, with coefficients
explicitly depending on the spatial coordinate $u$. In the small-amplitude
limit, the wave dynamics is governed by linear integral-differential equations.
It is using the conformal variables $u$ and $v$, 
that allows us to obtain these equations 
in exact form, contrary to the traditional approach where even linearized
equations can be obtained only approximately by expansion in the small
parameter, the slope of the bottom. The definition ``moderate'' 
for this variant emphasizes that straightening of the bottom without 
straightening the free boundary is able to provide not more 
than a weakly-nonlinear theory. Nevertheless, such a theory seems to be helpful
and applicable in many practical cases when wave amplitude is small. 
The results of this part 
of the work are the derivation of the Hamiltonian functional
for  weakly-nonlinear potential surface waves in canals having arbitrary bottom
shape, as well as calculations for band structure of  
spectrum for a number of periodic bottom profiles. As an example how to treat 
the linearized equations, also the problem is considered of wave reflection 
on a smooth ``step'' -- changing the depth from $h_1$ to $h_2$. 

The other variant of using the conformal mappings may be called ``radical''
in the sense it is valid for arbitrary shape of the bottom 
and for arbitrary shape of the free surface.
It is an exact combined theory where a  time-dependent conformal mapping
straightens both the bottom and the free boundary (Sec. 3). Such a mapping can 
be represented as the result of two mappings: $x+iy=z(\zeta(w,t))$, 
where the first function $\zeta(w,t)$ maps the horizontal stripe $-1<v<0$ 
onto the region $D_\zeta(t)$ with the straight lower boundary 
($\mbox{Im\,}\zeta=-1$)
and with a perturbed upper boundary, after that the time-independent function
$z(\zeta)$ maps the half-plane $\mbox{Im\,}\zeta>-1$ onto the region $y>-H(x)$
in the physical plane bounded  from below by the bottom. The shape of the free
surface will be described by the formula $X+iY=Z(u,t)=z(\zeta(u,t))$. 
However, it appears that exact nonlinear equations for $Z(u,t)$
in the inhomogeneous case have the same form as the known equations 
for waves over a horizontal bottom \cite{DZK96},
but with different analyticity requirements imposed on the solutions.
Numerical solutions obtained by the spectral method are presented that describe
a running and breaking wave (Sec. 4).

\section{Weakly-nonlinear theory}

So, suppose we know the analytical function $z(w)=x(u,v)+iy(u,v)$ which maps the
horizontal stripe $-1<\mbox{Im\,} w<0$ onto the region occupied by the fluid
at rest, and this function takes real values on the real axis: $z(u)=x(u,0)$.
The velocity field is irrotational, and the velocity potential $\varphi(u,v)$
satisfies the Laplace equation $\varphi_{uu}+\varphi_{vv}=0$ in the flow region
$-1<v<V(u,t)$, with the boundary conditions $\varphi_v|_{v=-1}=0$,
$\varphi|_{v=V(u)}=\psi(u)$. Due to conformal invariance of the Laplace
equation in 2D-space, hence equation $\varphi_{xx}+\varphi_{yy}=0$ is
satisfied as well, with no-penetration boundary condition on 
the bottom: $\partial\varphi/\partial n|_{y=-H(x)}=0$. Let us now take into
account the fact that the Lagrangian functional for potential surface 
waves has the following structure \cite{Z68,ZK97,DZK96}:
\begin{equation}\label{L_general_uv}
{\cal L}=\int\psi\dot\eta dx -{\cal H}
=\int\psi(Y_t X_u-Y_u X_t)du -{\cal H},
\end{equation}
where $Y(u,t)=y(u,V(u,t))$, $X(u,t)=x(u,V(u,t))$, 
and the Hamiltonian functional ${\cal H}$  is the total energy of the system --
sum of the kinetic energy and the potential energy in gravitational
field (in this paper we neglect surface tension effects, though they can be
easily incorporated by adding to the Hamiltonian the surface energy). In our
variables
\begin{eqnarray}
{\cal H}&=&\frac{1}{2}\int du\int\limits_{-1}^{V(u)}
(\varphi_u^2+ \varphi_v^2)d v \nonumber\\
\label{H_general}
&+& \frac{g}{2}\int y^2(u,V(u)) \frac{d}{du}x(u,V(u)) du.
\end{eqnarray}
This system has the obvious stable equilibrium $\psi=0$, $V=0$, hence
one may consider weak oscillations near this equilibrium state. In a standard
way (see, for instance \cite{ZK97}), 
let us expand the Lagrangian (\ref{L_general_uv}) in powers of the 
dynamical variables $\psi$ and $V$. 
It is clear that due to the symmetry principle the expansion for 
$y(u,v)$ contains only the odd powers of $v$, while  the expansion for $x(u,v)$
contains only the even powers of $v$. Therefore up to the third order in 
powers of $\psi$ and $V$ the Lagrangian (\ref{L_general_uv}) is equal to
\begin{equation}\label{L_3_uv}
\tilde{\cal L}=\int \psi V_t  x'^2(u)du
-{\cal K}^{(2)}\{\psi\}-{\cal K}^{(3)}\{\psi,V\}-{\cal P}^{(2)}\{V\},
\end{equation}
where $x'(u)=z'(u+0i)=x_u(u,0)$, and the equality $y_v=x_u$ has been 
taken into account in the first integral
in r.h.s. The expansion for the kinetic energy (calculation of the functionals
${\cal K}^{(2)}$ and ${\cal K}^{(3)}$) is performed in a standard manner 
\cite{ZK97,DZK96} and gives 
\begin{equation}\label{K_2}
{\cal K}^{(2)}\{\psi\}=\frac{1}{2}\int\psi[\hat k\tanh \hat k] \psi du,
\end{equation}
\begin{equation}\label{K_3}
{\cal K}^{(3)}\{\psi,V\}
=\frac{1}{2}\int [\psi_u^2-([\hat k\tanh \hat k] \psi)^2]V du.
\end{equation}
Here the linear Hermitian operator  $[\hat k\tanh \hat k]$ has been introduced, 
acting as
\begin{equation}\label{k_th_k}
[\hat k\tanh \hat k] \psi(u)=-\mbox{P.V.}\int\limits_{-\infty}^{+\infty}
\frac{\psi_{\tilde u}(\tilde u) d\tilde u}{2\sinh[(\pi/2)(\tilde u-u)]}.
\end{equation}
In Fourier-representation this operator simply multiplies the
Fourier-harmonics $\psi_k=\int \psi(u)\exp(-iku)du$ by $ k\tanh k$. Quadratic on $V$ part of the
potential energy is 
\begin{equation}\label{P_2}
{\cal P}^{(2)}\{V\}=\frac{g}{2}\int V^2 x'^3(u) du.
\end{equation}
It is convenient to deal with the function $\xi(u,t)=V(u,t) x'^2(u)$ 
canonically conjugated to $\psi(u,t)$, and  write the corresponding
up-to-third-order Hamiltonian in terms of $\xi$  and $\psi$:
\begin{eqnarray}
\tilde{\cal H}\{\xi,\psi\}&=&
\frac{1}{2}\int\psi[\hat k\tanh \hat k] \psi du+
\frac{g}{2}\int \frac{\xi^2}{ x'(u)} du\nonumber\\
\label{tilde_H}
&+&\frac{1}{2}\int 
\frac{[\psi_u^2-([\hat k\tanh \hat k] \psi)^2]\xi}{x'^2(u)} du.
\end{eqnarray}
Physically, this asymptotic expansion of the Hamiltonian is on a small
parameter -- the slope of the free surface (see \cite{ZK97} for more comments
and references).
The  weakly-nonlinear equations of motion have the standard Hamiltonian 
structure
\begin{eqnarray}
 \xi_t&=&\frac{\delta\tilde{\cal H}}{\delta\psi}=
[\hat k\tanh \hat k] \psi -\frac{\partial}{\partial u}\left(
\frac{\xi\psi_u}{x'^2(u)}\right)\nonumber\\
\label{xi_t_1}
&&\qquad-[\hat k\tanh \hat k]\left(
\frac{\xi [\hat k\tanh \hat k]\psi}{x'^2(u)}
\right) ,
\end{eqnarray} 
\begin{equation}\label{psi_t_1}
-\psi_t=\frac{\delta\tilde{\cal H}}{\delta\xi}=g\frac{\xi}{x'(u)}
+\frac{[\psi_u^2-([\hat k\tanh \hat k] \psi)^2]}{2x'^2(u)}.
\end{equation}

If $|x''(u)/x'(u)|\ll 1$, then $x'(u)$ is approximately 
equal to the equilibrium depth $H(u)$. For long waves over 
a such slowly varying bottom, only Fourier-harmonics $\psi_k$ and $\xi_k$ 
 with small $k$ are excited, so in this case the Hamiltonian (\ref{tilde_H}) 
can be simplified to the local form
\begin{equation}\label{tilde_H_long_waves}
\tilde{\cal H}_{\scriptsize \mbox{l}}\!=\!\!
\int\!\!\left[\frac{\psi_u^2}{2}-\frac{\psi_{uu}^2}{6}
+\frac{2\psi_{uuu}^2}{15}
+\frac{g\xi^2}{ 2x'(u)}
+\frac{\xi[\psi_u^2-\psi_{uu}^2]}{2x'^2(u)}\right] du,
\end{equation}
which is suitable for consideration of such phenomena as 
interaction of solitons with the bottom topography.

\subsection{Linearized equations}

Now let us consider the linearized system
\begin{equation}\label{lin_xi_psi}
\xi_t=[\hat k\tanh\hat k]\psi, \qquad
-\psi_t=g \frac{\xi}{x'(u)}.
\end{equation}
For a monochromatic wave ($\xi,\psi\propto \exp(-i\omega t)$) 
Eqs. (\ref{lin_xi_psi}) are reduced to the single integral equation
\begin{equation}\label{monochrom}
\left(\frac{\omega^2}{g}x'(u)-\hat k\tanh\hat k\right)\psi_\omega(u)=0.
\end{equation}
In the low-frequency limit this equation can be considerably simplified.
A variant of simplification is to introduce a new function $f$
by the equality $\hat k\tanh\hat k\psi_\omega(u)=-f_{uu}$. Then we 
obtain the equation 
\begin{equation}
\left(\frac{\omega^2}{g}x'(u)\hat k\coth\hat k
+(d/du)^2\right)f=0.
\end{equation}
The low-frequency limit corresponds to long wave-lengths, when 
$\hat k\coth\hat k\approx 1+\hat k^2/3=1-(1/3)(d/du)^2$, so we have to deal
with the second-order differential equation
\begin{equation}\label{SecondOrderAppr}
f_{uu}(u)+\frac{\frac{\omega^2}{g}x'(u)}{1-\frac{\omega^2x'(u)}{3g}}f(u)=0.
\end{equation}
where ${\omega^2x'(u)}/{g}$ should be small (only in this case the wave 
length is indeed effectively long;
remember that $x'(u)$ is of the same order as the depth).

Higher-order approximations to equation (\ref{monochrom}) can be derived in a
similar manner, for instance by change $\psi_\omega(u)=[\cosh\hat k] f(u)$ 
and subsequent expanding $[\cosh\hat k]$ and 
$[\hat k\sinh\hat k]$ in powers of $\hat k^2=-(d/du)^2$.

As an explicit example of using Eq.(\ref{SecondOrderAppr}), 
we consider reflection of a long wave from a 
step-shaped bottom inhomogeneity  described by the function 
\begin{equation}
z(w)=h_1w+\frac{(h_2-h_1)}{\alpha}\ln(1+e^{\alpha w}), 
\end{equation}
where  $h_1>h_2>0$, $0<\alpha \ll\pi$. If frequency of the wave is small,
$\omega\ll\sqrt{g/h_1}$, then equation (\ref{SecondOrderAppr}) may be 
applied. Calculating the derivative 
\begin{equation}\label{h1_h2}
z'(w)=h_1+(h_2-h_1)\frac{1}{1+e^{-\alpha w}}=
\frac{h_1e^{-\alpha w}+h_2}{e^{-\alpha w}+1},
\end{equation}
we have for $f(u)$ the equation
\begin{equation}\label{eq_for_f_step}
f_{uu}(u)+\frac{\omega^2}{g}
\frac{[\tilde h_1C e^{-\alpha u}+\tilde h_2]}{[C e^{-\alpha u}+1]}f(u)=0,
\end{equation}
where
\begin{equation}
\tilde h_1=\frac{h_1}{1-\frac{\omega^2h_1}{3g}},\quad
\tilde h_2=\frac{h_2}{1-\frac{\omega^2h_2}{3g}},\quad
C=\frac{1-\frac{\omega^2h_1}{3g}}{1-\frac{\omega^2h_2}{3g}}.
\end{equation}
A general solution for equation (\ref{eq_for_f_step}) is known \cite{LL3}.
In particular, the reflection coefficient is given by the expression
\begin{equation}
R(\omega)=\left(\frac{\sinh[\frac{\pi\omega}
{\alpha\sqrt{g}}(\sqrt{\tilde h_1}-\sqrt{\tilde h_2})]}
{\sinh[\frac{\pi\omega}
{\alpha\sqrt{g}}(\sqrt{\tilde h_1}+\sqrt{\tilde h_2})]}\right)^2.
\end{equation}

\subsection{Periodic bottom: The band structure of the spectrum}

\begin{figure}
\begin{center}
  \epsfig{file=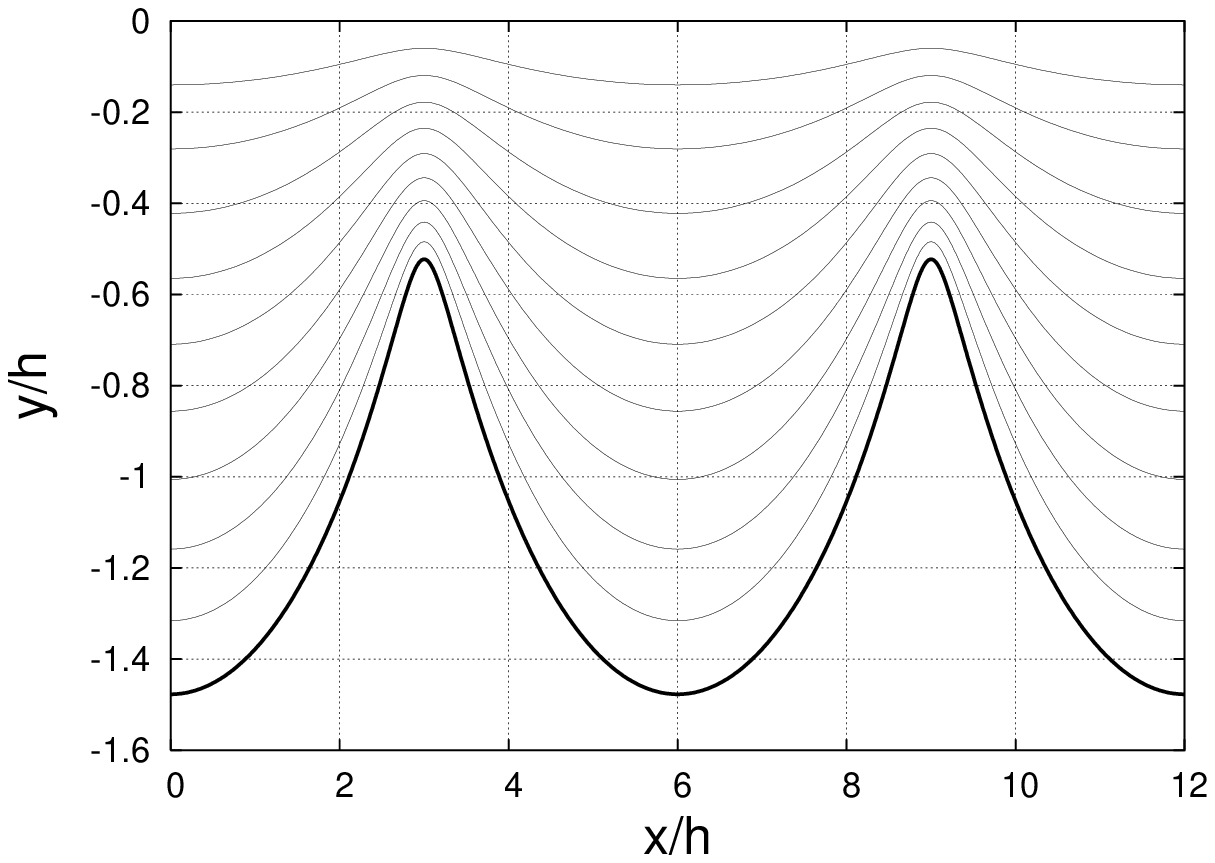,width=72mm}
  \epsfig{file=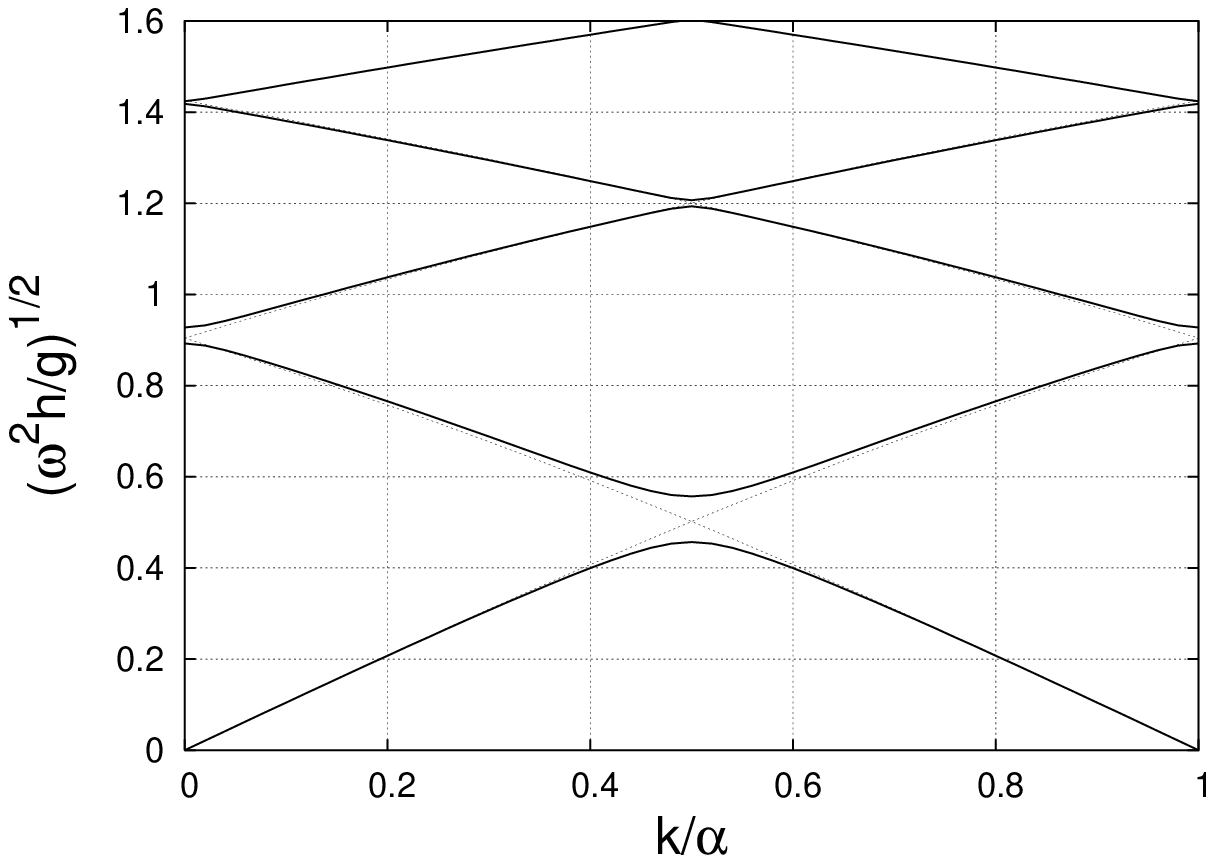,width=72mm}
\end{center}
\caption{\small Shape of the bottom, levels of constant $v$, 
and the band structure for  $z(w)/h=w+(2\epsilon/\alpha)\sin(\alpha w)$, 
with  $\alpha=\pi/3$, $\epsilon=0.2$.} 
\label{AnotherBandStructure}
\end{figure}

Interesting phenomena occur if shape of the bottom is periodic:
\begin{equation}
z'(w)=h\sum_n a_n\exp(in\alpha w),\qquad a_{-n}=\bar a_n.
\end{equation}
Here $h$ is a dimensional parameter, $a_n$ are some complex 
Fourier-coefficients. 
Obviously, $x'(u)=z'(u)>0$ and $|a_n|$ decay rapidly at large $|n|$, since
$z'(w)$ does  not have any singularities at $-1<\mbox{Im\,}w<1$.
The equation (\ref{monochrom}) for eigen-functions $\psi_\lambda(u)$ (where 
$\lambda=\omega^2h/g$) 
now has the form 
\begin{equation}\label{eq_periodic_u}
\lambda\left(\sum_n a_n\exp(in\alpha u)\right)\psi(u)-
[\hat k\tanh\hat k]\psi(u)=0,
\end{equation}
or in Fourier-representation
\begin{equation}
\lambda\sum_n a_n\psi_{k-n\alpha}=k\tanh k\,\, \psi_k.
\end{equation}
For convenience let us denote
\begin{equation}
F_\nu=\alpha\nu\,\tanh (\alpha\nu),\qquad \Psi_\nu=\psi_{\alpha\nu}.
\end{equation}
Now we have the infinite chain of linear equations
\begin{equation}\label{Psi_nu}
\lambda\sum_n a_n\Psi_{\nu-n}=F_\nu\Psi_\nu,
\end{equation}
where  $\Psi_{\nu_1}$ and $\Psi_{\nu_2}$  interact if the difference
between $\nu_1$ and $\nu_2$ is an integer number. Let us fix some $\nu$. 
Nontrivial solutions of the system (\ref{Psi_nu}) exist only
at definite values $\lambda=\lambda_m(\nu)$, where $m=1,2,3,\dots$. 
It is necessary to note that the functions $\lambda_m(\nu)$ are periodic:
$\lambda_m(\nu+1)=\lambda_m(\nu)$, and even: $\lambda_m(-\nu)=\lambda_m(\nu)$.
This determines the band structure of the spectrum with frequency gaps 
(see Figs.\ref{BandStructure-1}-\ref{AnotherBandStructure}).
For numerical computing $\lambda_m(\nu)$ it is necessary to cut the infinite
chain (\ref{Psi_nu}) at some large but finite length, thus considering only 
$\nu$ between $-N$ and $N$. Practically $N$ should be several times larger than 
the index $m$ of $\lambda_m$ . Numerical results for $\sqrt\lambda_m$ 
shown in the figures 
\ref{BandStructure-1}-\ref{AnotherBandStructure} were
obtained with the help of the mathematical package Maple 8 taking $N=10$. 

Fig.\ref{AnotherBandStructure} shows that in some cases even for strongly 
undulating bottom the coefficients $a_n$ with $n\ge 1$ can be still small
($a_1=\epsilon=0.2\ll 1$). 
In these cases it is easy to calculate
analytically in the main approximation positions of the gaps. 
For example, let us consider the bottom profile as in 
Fig.\ref{AnotherBandStructure}, where $x'(u)=h(1+2\epsilon\cos(\alpha u))$.  
The gaps in spectrum correspond to integer or half-integer $\nu$'s.
It is important that at these values of $\nu$, solutions of the linear chain
(\ref{Psi_nu}) possess definite parity, 
in  the sense that $\Psi_{-\nu}=\pm\Psi_{\nu}$.
This allows us in gap calculation consider only positive $\nu$. Let us
first consider half-integer $\nu$'s and the corresponding half-infinite chain
\begin{eqnarray}
(\lambda-F_{1/2})\Psi_{1/2}
+\lambda\epsilon(\pm \Psi_{1/2}+\Psi_{3/2})&=&0,\\
(\lambda-F_{3/2})\Psi_{3/2}+\lambda\epsilon(\Psi_{1/2}+\Psi_{5/2})&=&0,\\
(\lambda-F_{5/2})\Psi_{5/2}+\lambda\epsilon(\Psi_{3/2}+\Psi_{7/2})&=&0,\\
\dots\qquad \qquad \qquad&&\nonumber
\end{eqnarray}
Obviously, the even and odd cases result in different $\lambda$'s, 
and it is  this difference that determines the gaps in spectrum.
For main-order calculation of the first and third gaps, 
we cut this chain: $\Psi_{7/2}=0$, $\Psi_{9/2}=0$, and so on. 
Now we have to solve the equation for zeros of the determinant $3\times3$
\begin{eqnarray}
[\{\lambda(1\pm\epsilon) -F_{1/2}\}(\lambda-F_{3/2})-\lambda^2\epsilon^2]
(\lambda-F_{5/2})&&\nonumber\\
-\lambda^2\epsilon^2[\lambda(1\pm\epsilon) -F_{1/2}]=0.&&
\end{eqnarray}
First we take $\lambda=F_{1/2}+\Delta_1$, where  $\Delta_1$ is a small quantity
of the order  $\epsilon$. In the main order $\Delta_1 \pm \epsilon F_{1/2}=0$,
and this gives us the first gap: 
$F_{1/2}(1-\epsilon)<\lambda< F_{1/2}(1+\epsilon)$.

For the third gap we write $\lambda=F_{3/2}+\Delta_3$, where $\Delta_3$ 
is of order $\epsilon^2$. The equation for $\Delta_3$ with the third order
accuracy is 
\begin{eqnarray}
[(F_{3/2}(1\pm\epsilon)-F_{1/2})\Delta_3-\epsilon^2F^2_{3/2}](F_{3/2}-F_{5/2})&&
\nonumber\\
-\epsilon^2F^2_{3/2}(F_{3/2}(1\pm\epsilon)-F_{1/2})=0.&&
\end{eqnarray}
From here we find 
$$
\Delta_3=\epsilon^2F^2_{3/2}\left[\frac{1}{(F_{3/2}-F_{5/2})}
+\frac{1}{(F_{3/2}(1\pm\epsilon)-F_{1/2})}\right],
$$
where we may keep only the second- and third-order terms. This gives us the
position of the third gap $\lambda^{(3)}_-<\lambda<\lambda^{(3)}_+$:
\begin{eqnarray}
\lambda^{(3)}_{\pm}&=&F_{3/2}+\epsilon^2F^2_{3/2}
\left[\frac{1}{(F_{3/2}-F_{5/2})}
+\frac{1}{(F_{3/2}-F_{1/2})}\right]\nonumber\\
&&\pm \frac{\epsilon^3F^3_{3/2}}{(F_{3/2}-F_{1/2})^2}.
\end{eqnarray}

Analogously, the gaps at integer $\nu$'s can be considered. 
These are  determined by the system
\begin{eqnarray}
(\lambda-F_{0})\Psi_{0}+\lambda\epsilon(\pm \Psi_{1}+\Psi_{1})=0,&&\\
(\lambda-F_{1})\Psi_{1}+\lambda\epsilon(\Psi_{0}+\Psi_{2})=0,&&\\
(\lambda-F_{2})\Psi_{2}+\lambda\epsilon(\Psi_{1}+\Psi_{3})=0,&&\\
\dots\qquad \qquad \qquad&&\nonumber
\end{eqnarray}
For instance, position of the second gap in second order is
given by the formulas
\begin{equation}
\lambda^{(2)}_-=F_1-\frac{\epsilon^2F_1^2}{F_2-F_1},\quad
\lambda^{(2)}_+=F_1(1+2\epsilon^2)-\frac{\epsilon^2F_1^2}{F_2-F_1}.
\end{equation}

\section{Exact theory}

In exact nonlinear theory, the shape of free boundary is given in 
parametric form by a compound function $z(\zeta(u,t))$, where $z(\zeta)$
is a known function completely determined by the bottom shape [for example,  
$z(\zeta)=h(\sqrt{(\zeta+i)^2-(b/h)^2}-i)$ corresponds to a narrow vertical
barrier of the height $b$ at $x=0$ on the straight horizontal bottom with
the depth $y=-h$]. The unknown function $\zeta(w,t)$ should be analytical in the
stripe $-1<\mbox{Im\,} w<0$ and the combination $[\zeta(u-i,t)+i]$ 
should take real values. These conditions relate the real and the imaginary 
parts of $\zeta(u,t)$ at the real axis \cite{DZK96}: 
\begin{equation}\label{zeta_rho}
\zeta(u,t)=u+(1+i\hat R)\rho(u,t),
\end{equation}
where $\rho(u,t)$ is a real function, and the linear anti-Hermitian operator 
$\hat R$ is $i\tanh k$ in Fourier-representation. In $u$-representation
\begin{equation}\label{R=i_th_k}
\hat R \rho(u,t)=\mbox{P.V.}\int\limits_{-\infty}^{+\infty}
\frac{\rho(\tilde u,t) d\tilde u}{2\sinh[(\pi/2)(\tilde u-u)]}.
\end{equation}
The inverse operator $\hat R^{-1}=\hat T=-i\coth \hat k$ acts as 
\begin{equation}\label{T=-i_cth_k}
\hat T \rho(u,t)=\mbox{P.V.}\int\limits_{-\infty}^{+\infty}
\frac{\rho(\tilde u,t) d\tilde u}{1-\exp[\pi(\tilde u-u)]}.
\end{equation}
Note that the previously considered operator $[\hat k\tanh \hat k]$ is 
$-\hat R\partial_u$.
The kinetic energy functional is now exactly equal to the expression at the 
r. h. s. of Eq.(\ref{K_2}). The Lagrangian for $\psi(u,t)$ and $\zeta(u,t)$ 
is given by the formula
\begin{eqnarray}
&&{\cal L}_{\mbox{\scriptsize exact}}=\int |z'(\zeta)|^2
\left(\frac{\bar\zeta_u \zeta_t-\zeta_u \bar\zeta_t}{2i}\right)\psi du
\nonumber\\
\label{L_exact}
&&+\frac{1}{2}\int\psi \hat R \psi_u du
-\frac{g}{2}\int \{\mbox{Im\,}z(\zeta)\}^2 
\mbox{Re\,}(z'(\zeta)\zeta_u) du\nonumber\\
&&+\int\Lambda\left[\frac{\zeta-\bar\zeta}{2i}
-\hat R\left(\frac{\zeta+\bar\zeta}{2}-u\right)\right]du,
\end{eqnarray}
where the (real) Lagrangian indefinite multiplier $\Lambda(u,t)$ 
has been introduced in order
to take into account the analytical properties of the function $\zeta$ 
given by Eq.(\ref{zeta_rho}). From the above Lagrangian one can obtain the
equations of motion. Variation of the action 
$\int{\cal L}_{\mbox{\scriptsize exact}}dt$ by $\delta\psi(u,t)$ gives the
equation
\begin{equation}\label{var_L_exact_psi}
|z'(\zeta)|^2(\bar\zeta_u \zeta_t-\zeta_u \bar\zeta_t)/(2i)
=- \hat R \psi_u,
\end{equation}
which can be easily transformed to the form (compare with \cite{DZK96})
\begin{equation}\label{kinematic_exact}
\zeta_t=-\zeta_u (\hat T +i)\left[
\frac{\hat R \psi_u}{|z'(\zeta)|^2|\zeta_u|^2}\right].
\end{equation}
The variation of the action by $\delta\zeta(u,t)$ 
results after simplifying in the equation
\begin{equation}\label{d_L_exact_d_zeta}
|z'(\zeta)|^2\left\{\psi_t\bar\zeta_u-\psi_u\bar\zeta_t
+g\bar\zeta_u\mbox{Im\,}z(\zeta)\right\}-(1+i\hat R)\Lambda=0.
\end{equation}
Since the product $\zeta_u(1+i\hat R)\Lambda$ has the same analytical
properties as both $\zeta_u$ and $(1+i\hat R)\Lambda$, 
we can multiply Eq.(\ref{d_L_exact_d_zeta}) by $\zeta_u$ and write
\begin{equation}\label{d_L_exact_d_zeta_tilde}
|z'(\zeta)|^2\left\{[\psi_t+g\mbox{Im\,}z(\zeta)]|\zeta_u|^2
-\psi_u\bar\zeta_t\zeta_u\right\}-(1+i\hat R)\tilde\Lambda=0,
\end{equation}
where $\tilde\Lambda$ is another real function.
The imaginary part of the above equation together with 
Eq.(\ref{var_L_exact_psi}) result in 
\begin{equation}
\tilde\Lambda=-\hat T[\psi_u\hat R\psi_u].
\end{equation}
Using this equality, we can reduce the real part of 
Eq.(\ref{d_L_exact_d_zeta_tilde}) to the form
\begin{equation}\label{Bernoulli_exact}
\psi_t+g\mbox{Im\,}z(\zeta)=-\psi_u\hat T\left[
\frac{\hat R \psi_u}{|z'(\zeta)|^2|\zeta_u|^2}\right]
-\frac{\hat T[\psi_u\hat R\psi_u]}{|z'(\zeta)|^2|\zeta_u|^2},
\end{equation}
which is the Bernoulli equation in the conformal variables.
Exact equations (\ref{kinematic_exact}) and (\ref{Bernoulli_exact}) 
[with given analytical function $z(\zeta)$ and with the condition
(\ref{zeta_rho})] completely determine the evolution 
of gravitational surface waves over the undulating bottom 
parameterized by a real parameter $r$ as 
$X_b(r)+iY_b(r)=z(\zeta)|_{\zeta=r-i}$. 

Equations (\ref{kinematic_exact}) and (\ref{Bernoulli_exact})
can be represented in another form by using the identity
$2\hat T[\psi_u\hat R\psi_u]=\psi_u^2-(\hat R\psi_u)^2$ \cite{DZK96}
and introducing the complex potential 
\begin{equation}\label{Phi_definition}
\Phi(u,t)=(1+i\hat R)\psi(u,t)
\end{equation}
(which is analytically continued to the stripe $-1<\mbox{Im\,}w<0$):
\begin{eqnarray}
\label{kinematic_exact_Phi_zeta}
\zeta_t&=&-\zeta_u (\hat T +i)\left[\frac{\mbox{Im\,}\Phi_u}
{|z'(\zeta)|^2|\zeta_u|^2}\right],\\
\Phi_t&=&-\Phi_u(\hat T+i)\left[
\frac{\mbox{Im\,}\Phi_u}{|z'(\zeta)|^2|\zeta_u|^2}\right]\nonumber\\
\label{Bernoulli_exact_Phi_zeta}
&&-(1+i\hat R)\left[\frac{|\Phi_u|^2}{2 |z'(\zeta)|^2|\zeta_u|^2}
+g\mbox{Im\,}z(\zeta)\right].
\end{eqnarray}
A very interesting point is that one can re-write equations 
(\ref{kinematic_exact}) and (\ref{Bernoulli_exact})
without the intermediate function $\zeta(u,t)$, but directly for $z(u,t)$.
Indeed, after multiplying Eq.(\ref{kinematic_exact}) by $z'(\zeta)$ we obtain 
the equations
\begin{equation}\label{kinematic_exact_z}
z_t=-z_u (\hat T +i)\left[
\frac{\hat R \psi_u}{|z_u|^2}\right],
\end{equation}
\begin{equation}\label{Bernoulli_exact_z}
\psi_t+g\mbox{Im\,}z=-\psi_u\hat T\left[
\frac{\hat R \psi_u}{|z_u|^2}\right]
-\frac{\hat T[\psi_u\hat R\psi_u]}{|z_u|^2},
\end{equation}
that is exactly the same system as was derived in \cite{DZK96} 
for a straight horizontal bottom. However, in our case analytical 
properties of the function $z(w,t)$ are different:  
\begin{equation}
\mbox{Im\,}z(u)\not=\hat R [\mbox{Re\,}(z(u)-u)].
\end{equation}
The only requirements for $z(w,t)$ now are that it should be analytical in the 
stripe $-1<\mbox{Im\,}w<0$ and the corresponding mapping should have a 
physical sense (no self-intersections are allowed).
The question may arise: Where is the bottom shape in 
Eqs. (\ref{kinematic_exact_z})-(\ref{Bernoulli_exact_z})? The answer is simple:
The shape of the bottom is an integral of motion for this system. Roughly
speaking, each particular solution of 
Eqs.(\ref{kinematic_exact_z})-(\ref{Bernoulli_exact_z}) corresponds to 
a flow over a definite topography determined by the initial condition
$z(r-i,0)$.

Analogously, 
Eqs.(\ref{kinematic_exact_Phi_zeta})-(\ref{Bernoulli_exact_Phi_zeta})
can be represented as 
\begin{equation}\label{kinematic_exact_Phi_z}
z_t=-z_u (\hat T +i)\left[\frac{\mbox{Im\,}\Phi_u}{|z_u|^2}\right],
\end{equation}
\begin{equation}\label{Bernoulli_exact_Phi_z}
\Phi_t=-\Phi_u(\hat T+i)\left[
\frac{\mbox{Im\,}\Phi_u}{|z_u|^2}\right]
-(1+i\hat R)\left[\frac{|\Phi_u|^2}{2|z_u|^2}+g\mbox{Im\,}z\right].
\end{equation}

\section{Numerical experiments}
\subsection{Different forms of equations}

For numerical simulations, still other equivalent forms of exact 
equations may be useful, since numerical stability depends dramatically on the
choice of dynamical variables \cite{D2001,ZDV2002,DZ2004,ZL2004}. 
Two alternative sets of equations were used in computations 
presented below. First, as it was pointed in \cite{D2001} for the case
of deep water, a good practical choice for the dynamical variables is $A=1/z_u$ 
and $B=\Phi_u/z_u$. It is easy to derive the equations of motion for $A(u,t)$ 
and $B(u,t)$  from 
Eqs.(\ref{kinematic_exact_Phi_z})-(\ref{Bernoulli_exact_Phi_z}),
and they are very elegant (compare with \cite{D2001}):
\begin{equation}\label{kinematic_exact_AB}
A_t=-A_u (\hat T +i)\mbox{Im\,}(B\bar A)
+A(\hat T +i)\partial_u\mbox{Im\,}(B\bar A),
\end{equation}
\begin{equation}\label{Bernoulli_exact_AB}
B_t=-B_u(\hat T+i)\mbox{Im\,}(B\bar A)
-A(1+i\hat R)\left[\partial_u\frac{|B|^2}{2}+g\mbox{Im\,}\frac{1}{A}\right].
\end{equation}
The variables $A$ and $B$ do allow stable numerical simulations for waves over 
varying seabed. However,  analytical properties of $A$ and $B$ are not 
restricted by conditions similar to Eq.(\ref{zeta_rho}),
and therefore the shape of the bottom is preserved in this case only 
approximately. 

The second set of variables, that were used in numerical experiment,
consists of two complex functions: 
$\zeta(u,t)$ and 
\begin{equation}
\beta(u,t)=\Phi_u(u,t)/\zeta_u(u,t),
\end{equation}
both having effectively controlled analytical properties. 
With this choice, the bottom shape is preserved exactly, 
but the corresponding equations of motion are slightly less compact:
\begin{eqnarray}
\label{kinematic_exact_zeta_beta}
\zeta_t&=&-\zeta_u (\hat T +i)
\mbox{Im\,}\left(\frac{\beta}{|z'(\zeta)|^2\bar \zeta_u}\right),\\
\beta_t&=&-\beta_u(\hat T+i)
\mbox{Im\,}\left(\frac{\beta}{|z'(\zeta)|^2\bar \zeta_u}\right)
\label{Bernoulli_exact_zeta_beta}\nonumber\\
&&-\zeta_u^{-1}(1+i\hat R)\partial_u
\left[\frac{|\beta|^2}{2|z'(\zeta)|^2}+g\,\mbox{Im\,}z(\zeta)\right].
\end{eqnarray}

It is necessary to explain here some important  details about space-periodic 
solutions of the system 
(\ref{kinematic_exact_zeta_beta})-(\ref{Bernoulli_exact_zeta_beta}), since
spectral numerical methods deal  with periodic functions.
Such solutions exist if the function $z'(\zeta)$ is periodic with a fixed
real period $L$, so that $z(\zeta+L)=L+z(\zeta)$. 
However, this does not imply that the functions $\zeta_u(u,t)$
and $\beta(u,t)$ have a fixed $u$-period. It would be so, 
but the linear operator $\hat T$ is singular at small $k$, and its action
on a constant function is not periodic in $u$: $\hat T C=Cu$. Thus 
in the right-hand-sides of the 
Eqs.(\ref{kinematic_exact_zeta_beta})-(\ref{Bernoulli_exact_zeta_beta})
we have non-periodic terms. Therefore $\zeta_t(u,t)$ and $\beta_t(u,t)$
cannot retain a constant $u$-period. However,
$\zeta_t(u,t)$ and $\beta_t(u,t)$ can be space-periodic with a 
{\bf time-dependent $u$-period}. So,
at arbitrary moment of time we will have the equality
$\zeta(u,t)+i=(L/2\pi)\zeta_*\left(2\pi\alpha(t)u/L,t\right)$, where
\begin{eqnarray}
\zeta_*(\vartheta,t)&=&\vartheta+i\alpha(t)
+\sum_{m=-\infty}^{+\infty}\frac{2\rho_m(t)\exp(im\vartheta)}
{1+\exp(2m\alpha(t))}\nonumber\\
&=&\vartheta+i\alpha(t)+(1+i\hat{\mathsf R}_\alpha)\rho(\vartheta,t),  
\end{eqnarray}
with an unknown real function 
$\alpha(t)$.
The unknown complex Fourier coefficients  $\rho_m(t)$ correspond
to a real ($2\pi$-periodic on the variable $\vartheta$) function 
$\rho(\vartheta,t)$:
$$
\rho(\vartheta,t)=
\sum_{m=-\infty}^{+\infty}\rho_m(t)\exp(im\vartheta),
\qquad \rho_{-m}(t)=\bar\rho_m(t).
$$
The linear operator $\hat{\mathsf R}_\alpha$  is diagonal in the discrete
Fourier representation:
${\mathsf R}_\alpha(m)=i\tanh(\alpha m)$.  

Analogously, $\beta(u,t)$ can be represented as 
$\beta=(gL/(2\pi))^{1/2}\beta_*(\vartheta,t)$, where
\begin{equation}
\beta_*(\vartheta,t)=\sum_{m=-\infty}^{+\infty}
\frac{2\chi_m(t)\exp(im\vartheta)}
{1+\exp(2m\alpha(t))}=(1+i\hat{\mathsf R}_\alpha)\chi(\vartheta,t).
\end{equation}
Equations of motion for the real functions  $\alpha(t)$, $\rho(\vartheta,t)$, 
and $\chi(\vartheta,t)$ follow from 
Eqs.(\ref{kinematic_exact_zeta_beta})-(\ref{Bernoulli_exact_zeta_beta}):
\begin{eqnarray}
\dot\alpha(t)\!\!&=&\!\!\frac{1}{2\pi}\int_0^{2\pi}\mbox{Im}
\left(\frac{-\beta_*(\vartheta)}
{|z_*'(\zeta_*)|^2\bar \zeta_*'(\vartheta)}\right)
d\vartheta,\label{dot_alpha}\\
\dot \rho(\vartheta,t)\!\!&=&\!\!\mbox{Re}\left(\zeta_*'(\hat{\mathsf T}_\alpha+i)
\mbox{Im}\left(\frac{-\beta_*(\vartheta)}
{|z_*'(\zeta_*)|^2\bar \zeta_*'(\vartheta)}\right)\right),
\label{dot_rho}\\
\dot \chi(\vartheta,t)\!\!&=&\!\!\mbox{Re}\Big(\beta_*'(\hat{\mathsf T}_\alpha+i)
\mbox{Im}\left(\frac{-\beta_*(\vartheta)}
{|z_*'(\zeta_*)|^2\bar \zeta_*'(\vartheta)}\right)\label{dot_chi}\nonumber\\
&-&\!\!\frac{1}{\zeta_*'}(1+i\hat{\mathsf R}_\alpha)\partial_{\vartheta}
\left[\frac{|\beta_*|^2}{2|z_*'(\zeta_*)|^2}+\mbox{Im}z_*(\zeta_*)\right]
\Big),
\end{eqnarray}
where  $z_*(\zeta_*)=(2\pi/L)z(-i+L\zeta_*/(2\pi))$, 
$\zeta_*'=\partial_{\vartheta}\zeta_*$ and so on.
The linear operator $\hat{\mathsf T}_\alpha$ is regular. In the discrete
Fourier representation it is defined as follows:
\begin{eqnarray}
{\mathsf T}_\alpha(m)&=&-i\coth(\alpha m), \qquad m\not=0;\nonumber\\
&=& 0,\qquad\qquad \qquad\quad m=0.
\end{eqnarray}
 
Two numerical experiments are briefly reported below, first of them employing
Eqs.(\ref{kinematic_exact_AB})-(\ref{Bernoulli_exact_AB}), and the second one
employing Eqs.(\ref{kinematic_exact_zeta_beta})-(\ref{Bernoulli_exact_zeta_beta}). 
Both the systems (\ref{kinematic_exact_AB})-(\ref{Bernoulli_exact_AB})
and  (\ref{kinematic_exact_zeta_beta})-(\ref{Bernoulli_exact_zeta_beta}) are 
equally convenient for numerical solution by spectral methods, 
inasmuch as the multiplications 
can be performed in $u$-representation while the linear operators $\hat R$ and
$\hat T$ (also the $u$-differentiation) are simple in the Fourier representation.
Efficient subroutine libraries for the fast Fourier transform are now
available. 
The integration schemes in both cases were based on the Runge-Kutta 
4-th order method, similarly to work \cite{ZDV2002}. 
For computing the discrete Fourier transform,
the FFTW library was used \cite{fftw3}. 
The length scale was normalized by a factor 
$h=L/(2\pi)$ and the velocity scale by $(gh)^{1/2}$.

\subsection{Numerical results for 
Eqs.(\ref{kinematic_exact_AB})-(\ref{Bernoulli_exact_AB})}
\begin{figure}
\begin{center}
  \epsfig{file=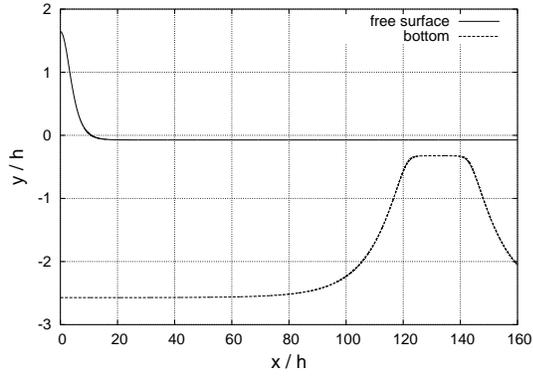,width=72mm} 
\end{center}
\caption{\small (i) Free surface and bottom for $t=0$. 
Only part of the entire periodic domain is shown.} 
\label{t0-a}
\end{figure}
\begin{figure}
\begin{center}
  \epsfig{file=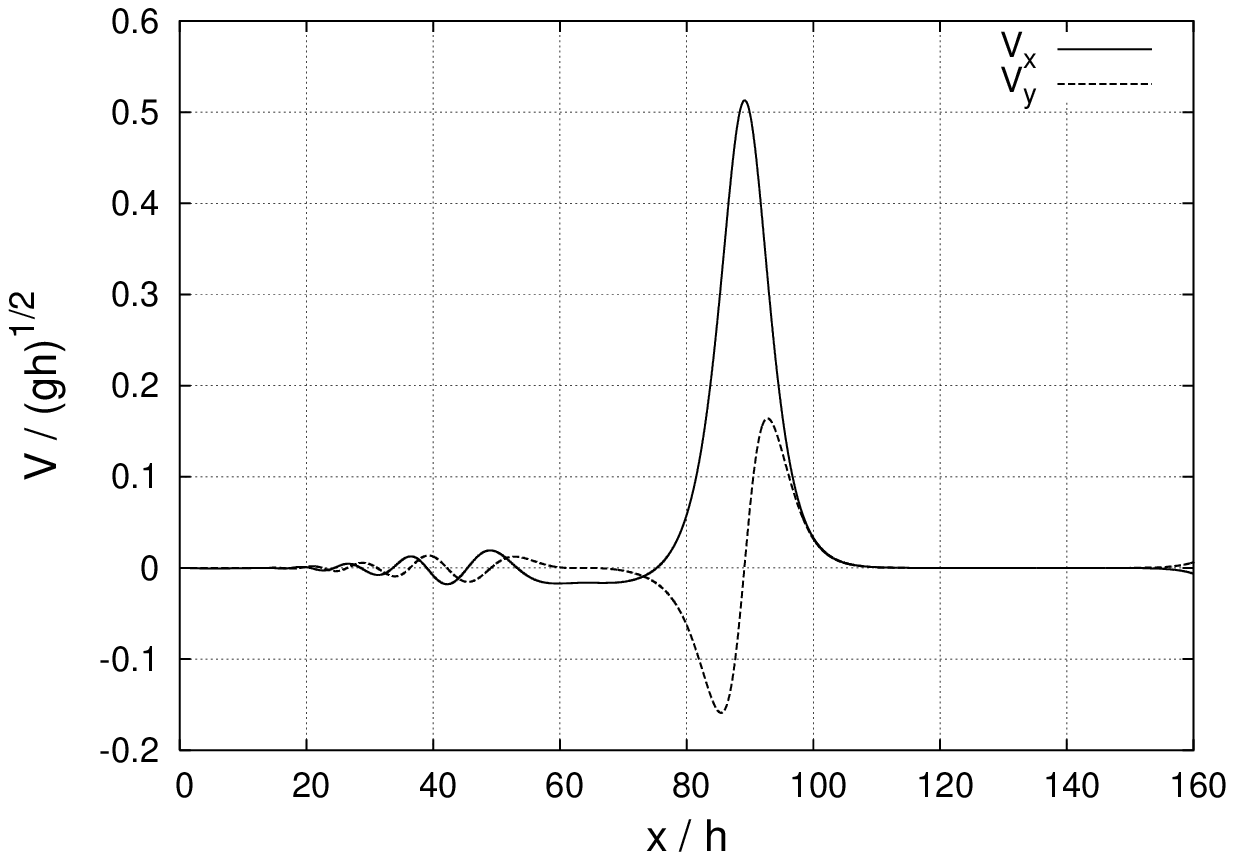,width=72mm}
  \epsfig{file=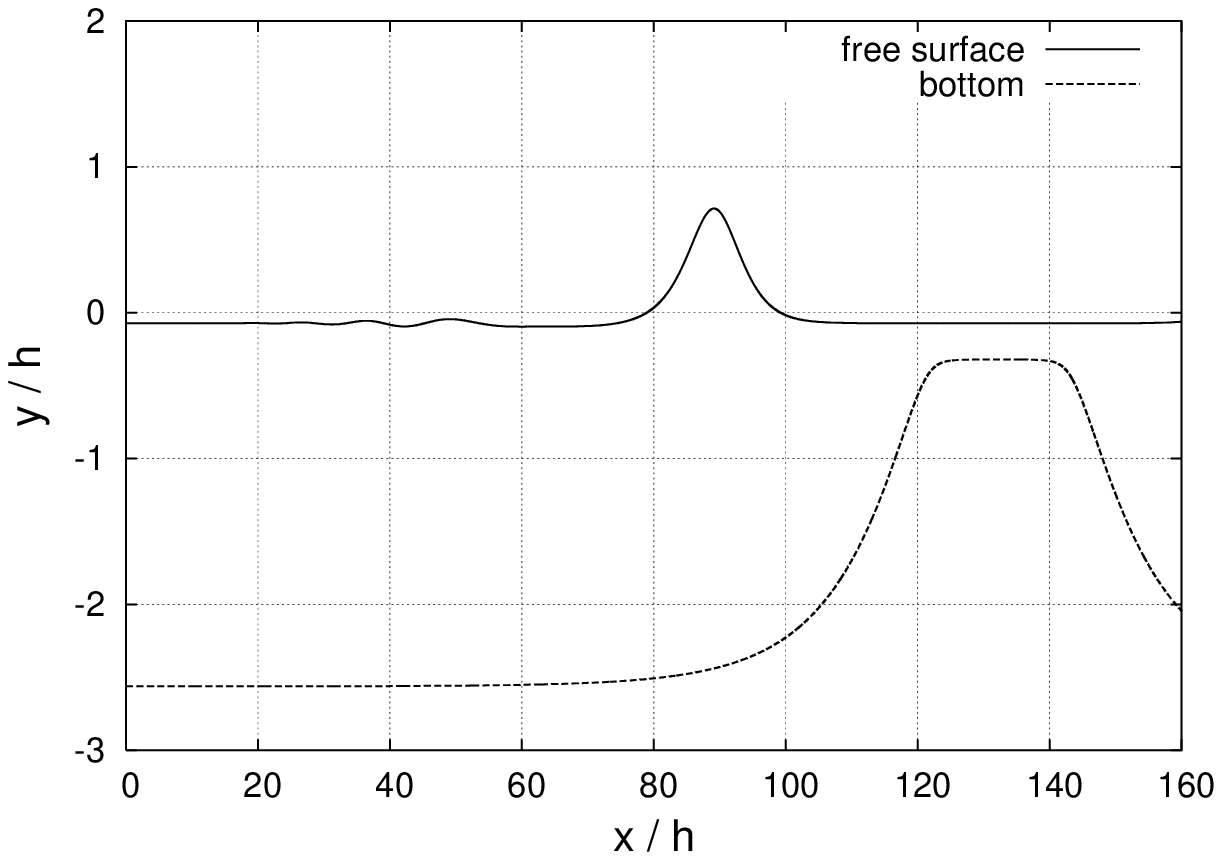,width=72mm}
\end{center}
\caption{\small (i) $t=50$: propagation stage.} 
\label{t1-a}
\end{figure}
\begin{figure}
\begin{center}
  \epsfig{file=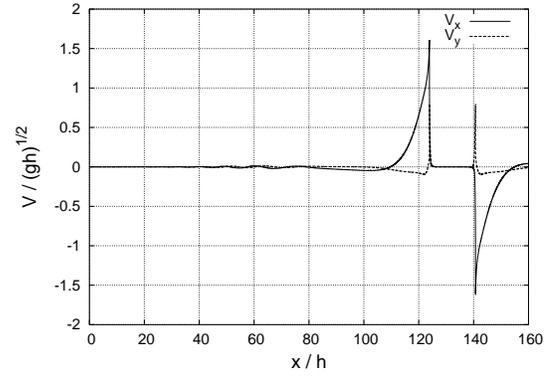,width=72mm}
  \epsfig{file=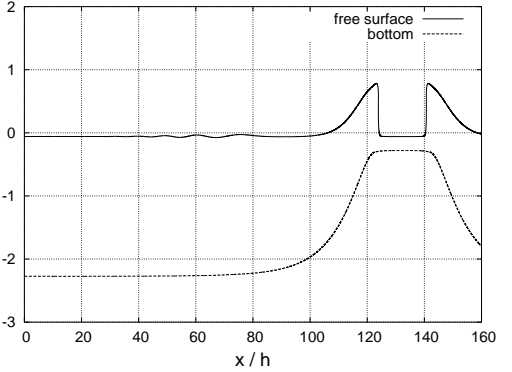,width=72mm}  
\end{center}
\caption{\small (i) $t=69$: breaking of the wave.} 
\label{t7-a}
\end{figure}
\begin{figure}
\begin{center}
  \epsfig{file=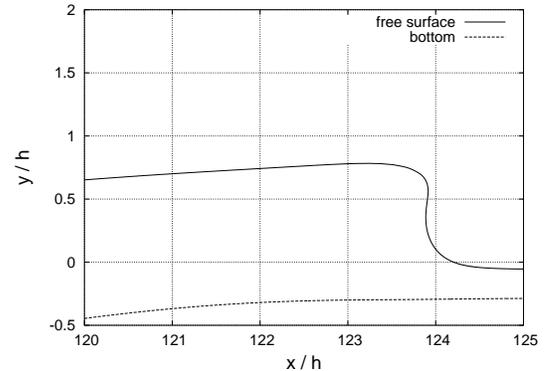,width=72mm} 
\end{center}
\caption{\small (i) Crest of the wave for $t=69$.} 
\label{crest7-a}
\end{figure}
The above remark about time-dependent $u$-period equally concerns the system 
(\ref{kinematic_exact_AB})-(\ref{Bernoulli_exact_AB}) as 
(\ref{kinematic_exact_zeta_beta})-(\ref{Bernoulli_exact_zeta_beta}). However,
for localized disturbances the relative change of $\alpha(t)$ remains small. 
Therefore solitary waves are possible to simulate with a constant $\alpha$
and with the regularized $\hat T$.
It was done  so the first numerical experiment [referred as (i)], 
where periodic boundary conditions were applied with the fixed
$u$-period $L=200$. The initial conditions were taken $B(u,0)=0$ and
$$
\frac{1}{A(u,0)}=z_u(u,0)=
\frac{2.5 E(u)+0.25}{E(u)+1}+\frac{1.4}{C(u)},
$$
where 
\begin{eqnarray}
E(u)&=&\exp[8.0 (\cos(2\pi u/L)-0.1)],\nonumber\\
C(u)&=&\cosh[(L/\pi)\sin\{\pi(u+i)/L\}].\nonumber
\end{eqnarray}
This initial configuration is symmetric (even). It results after some time in 
two oppositely propagating, nearly solitary waves. The waves are created in the
region where the depth is maximal.  In the course of motion each wave approaches
a shallow region where the surface profile $y=\eta(x,t)$ becomes steeper  and finally 
multi-valued. 

Some of the results of this numerical experiment are presented in  
Figs.\ref{t0-a}-\ref{crest7-a}, where the  velocity distribution on the surface is 
shown, as well as shapes of free surface and of the bottom for several moments 
of time. In general, the computed wave profiles look quite realistic,
though the present theory does not take into account viscous effects.
Steeping of the wave profile is clearly seen. It should be noted, 
however, in these simulations the bottom 
shape is preserved only  approximately due to discretization errors, 
and the same concerns the total energy. The computation was stopped well before
the moment of formation of a singularity on the crests of the waves, 
when the numerical scheme becomes invalid. 
In real world this moment corresponds to development of
a three-dimensional instability resulting in vortices, splash and foam.

\subsection{Numerical results for 
Eqs.(\ref{kinematic_exact_zeta_beta})-(\ref{Bernoulli_exact_zeta_beta})}

\begin{figure}
\begin{center}
  \epsfig{file=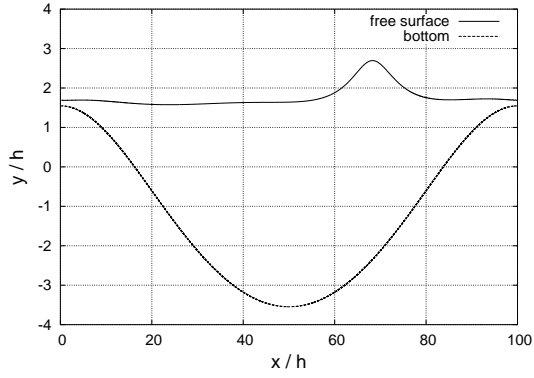,width=72mm}   
\end{center}
\caption{\small (ii) The bottom profile and shape of the free surface 
at $t=0$. The velocity field is everywhere zero.} 
\label{t0}
\end{figure}
\begin{figure}
\begin{center}
  \epsfig{file=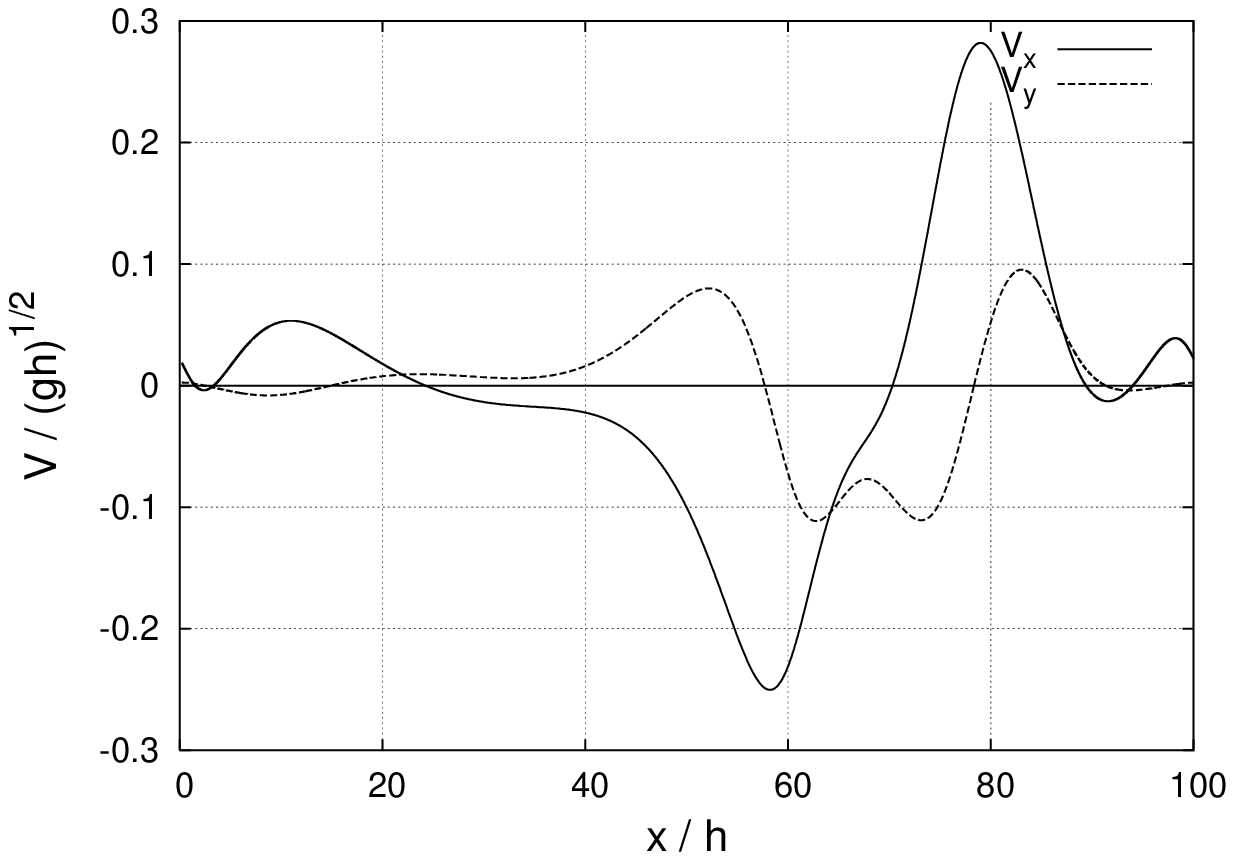,width=72mm}
  \epsfig{file=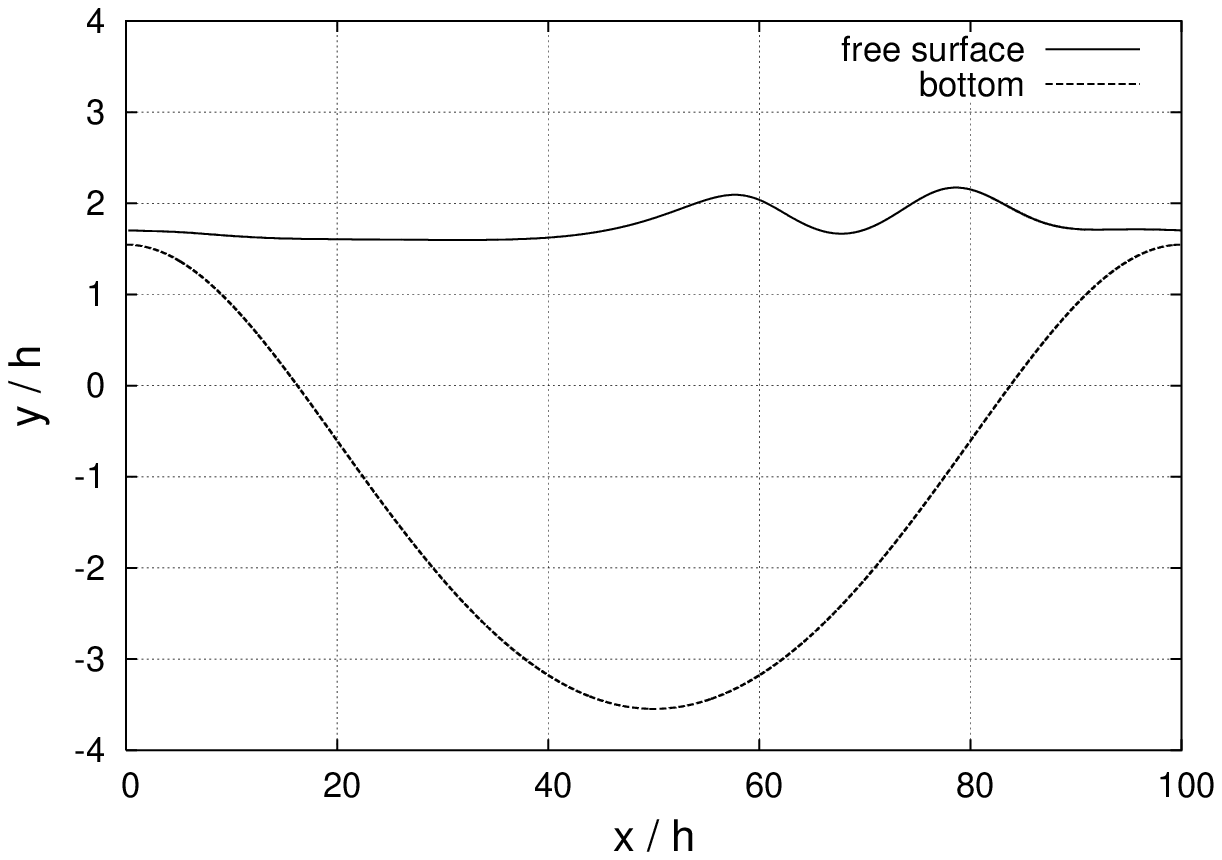,width=72mm}
\end{center}
\caption{\small (ii) Velocity distribution on free surface and shape
of the surface at $t=6$.} 
\label{t2}
\end{figure}

\begin{figure}
\begin{center}
  \epsfig{file=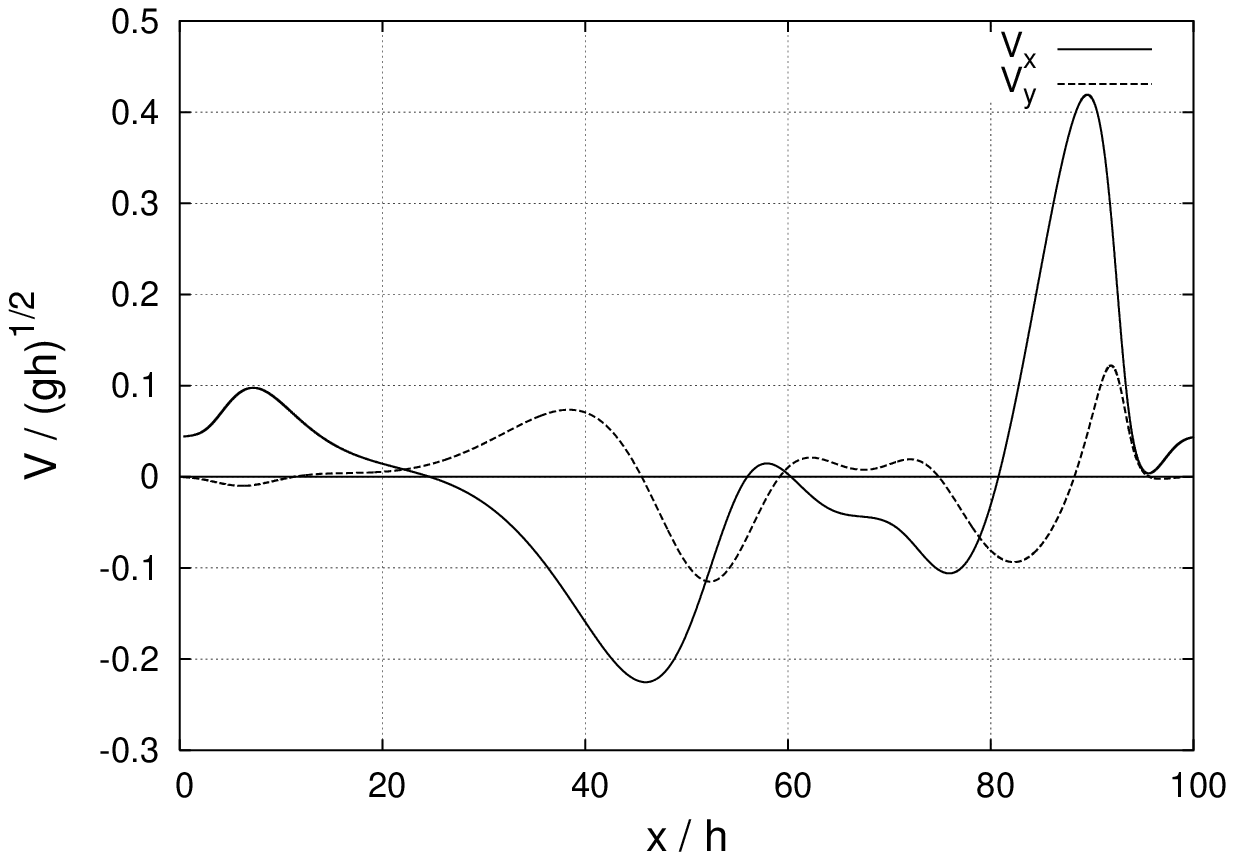,width=72mm}
  \epsfig{file=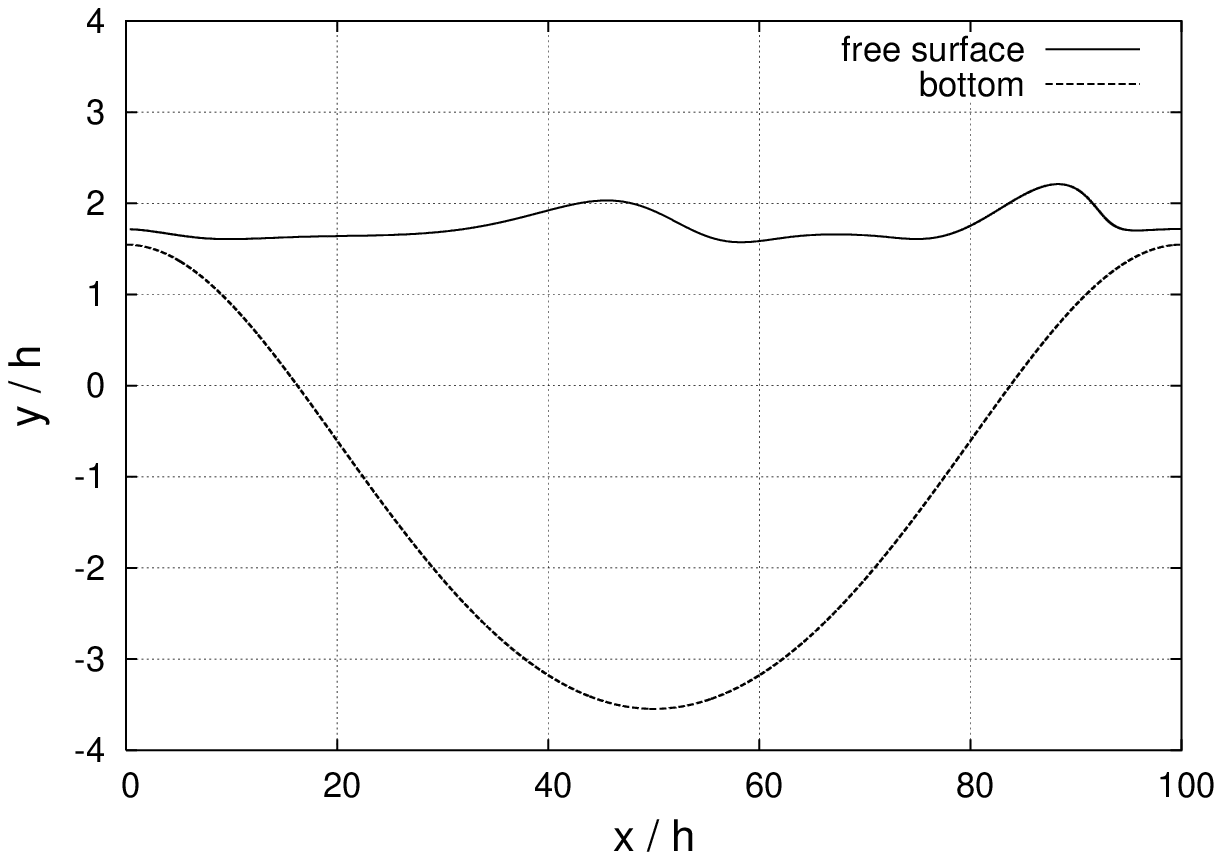,width=72mm}
\end{center}
\caption{\small (ii) The same as in Fig.\ref{t2}, at $t=12$.} 
\label{t4}
\end{figure}
\begin{figure}
\begin{center}
  \epsfig{file=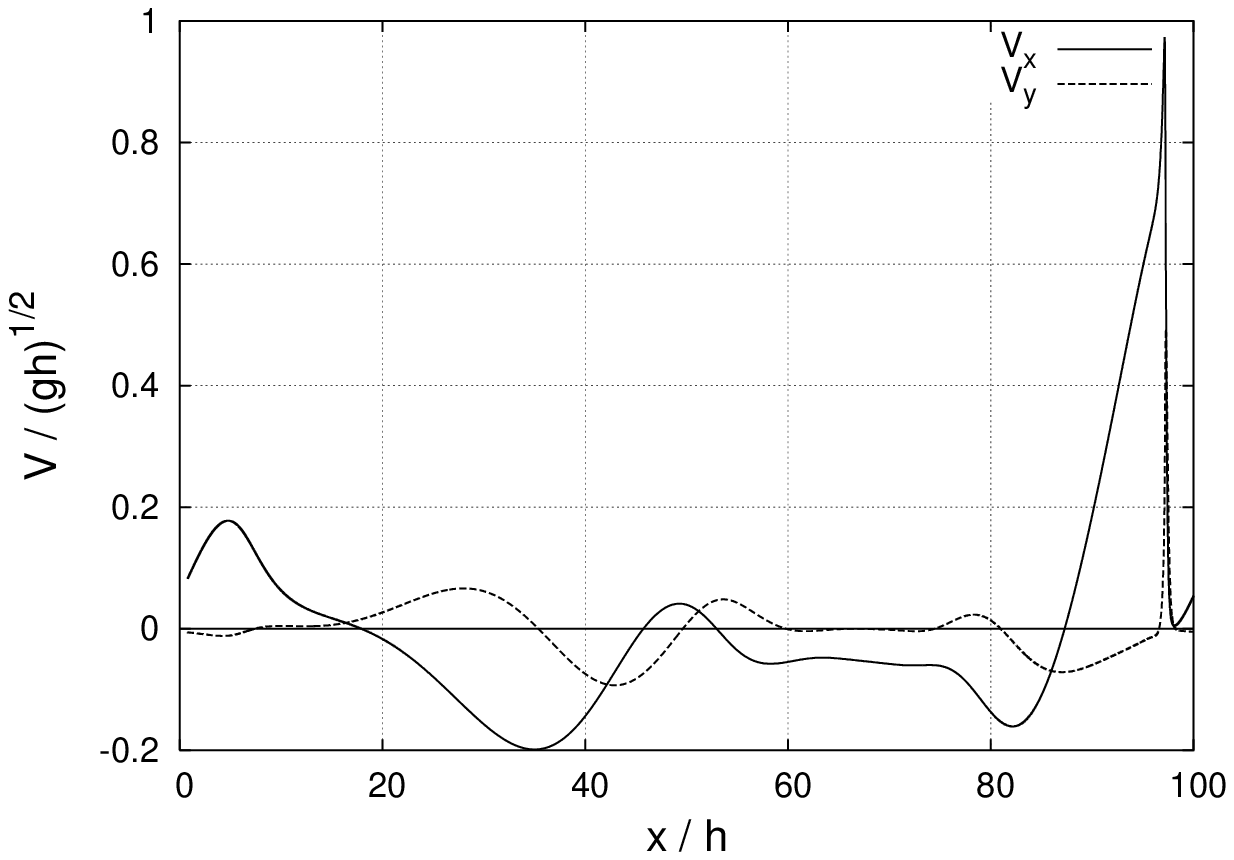,width=72mm}
  \epsfig{file=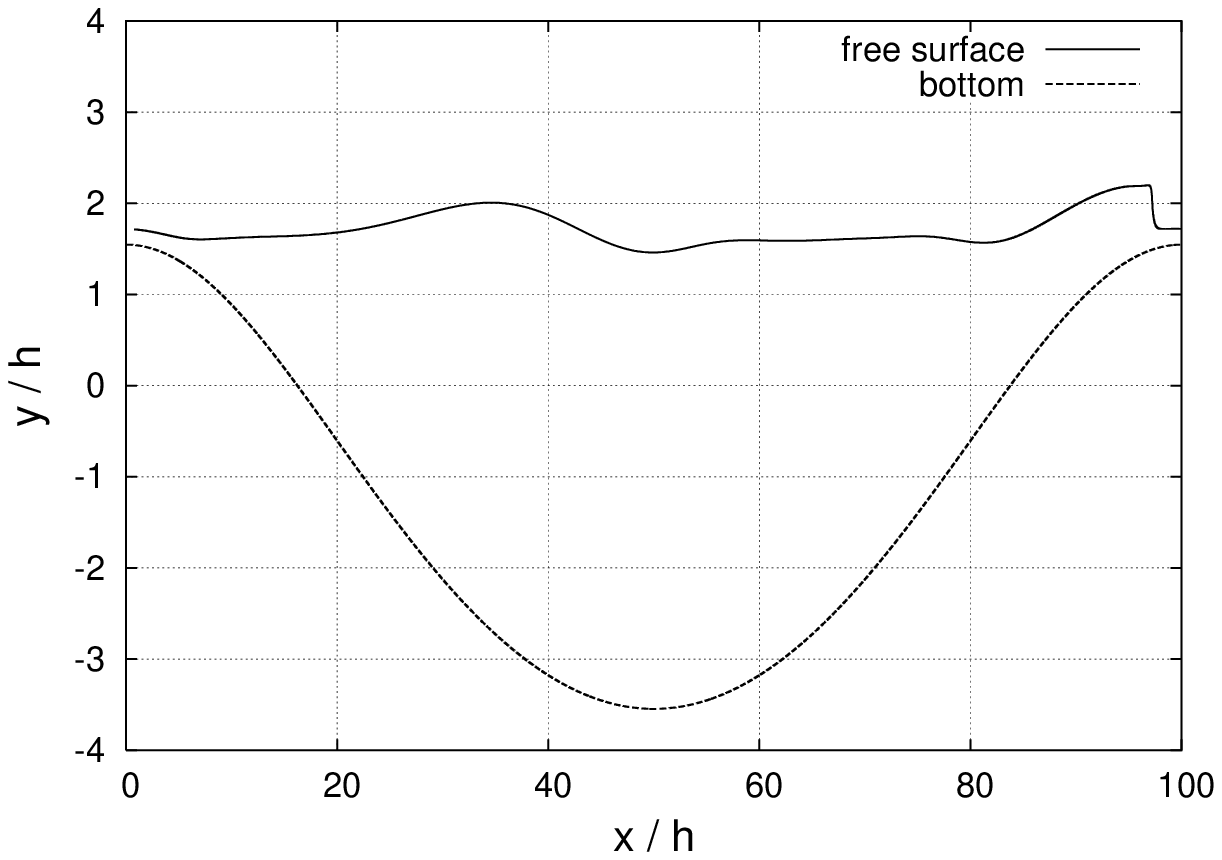,width=72mm}   
\end{center}
\caption{\small (ii) The same as in Fig.\ref{t2}, at $t=17$.} 
\label{t7}
\end{figure}

In the second numerical experiment [referred as (ii)], the shape of the bottom 
was fixed by analytical function
$$
z(\zeta)=\zeta+i\frac{Ld}{2\pi}\exp(2\pi i(\zeta+i)/L),
$$
with the dimensionless parameters $L=100$, $d=0.16$. 
The initial velocity field was taken zero: $\beta(u,0)=0$, 
while $\zeta(u,0)$ had the form
$$
\zeta(u,0)=u+\Theta(0.63,\, 0.0, \,u+i) +0.06\,\, \Theta(0.9,\, 0.04,\, u+i),
$$
where
$$
\Theta(r, p, w)\equiv-i\frac{L}{2\pi}\ln\left(
\frac{1+r\exp(-2\pi i(w-pL)/L)}{1+r\exp(+2\pi i(w-pL)/L)}\right).
$$
Qualitatively, these initial conditions are similar to those in the first
experiment, however now two oppositely propagating waves are created over 
inclined region of the bottom, 
so there is no left-right symmetry in the evolution.

The full system (\ref{dot_alpha})-(\ref{dot_chi}) was solved with a 
high accuracy for $m$ in the limits $-7000<m<7000$
(the energy conservation was up to 12 digits for a long ``smooth''
initial  stage of the evolution, and it was up to 5 digits for the final 
stage just before the breaking). This more accurate numerical solution was 
compared to a less accurate solution obtained with fixed $\alpha=\alpha(0)$. 
The difference was found very small.

The corresponding numerical results are presented in Figs.\ref{t0}-\ref{t7}.
Here we again observe steeping of the wave profile with the tendency towards
finite time singularity formation on the crest. Such behavior indeed takes
place in natural conditions when the flows are almost two-dimensional.

\section{Summary}

In this paper we have derived approximate weakly-nonlinear, as well as exact
nonlinear equations of motion for potential water waves over a strongly 
inhomogeneous bottom. The consideration was based on using the conformal 
mappings. For linear waves over periodic seabed, 
the band structure of the spectrum has been calculated.

Though the obtained exact equations can be written in formally the same form
as those derived in \cite{DZK96} for a straight horizontal bottom,  
but admissible solutions have different analytical properties if the bottom 
is inhomogeneous. When the equations are written in this form, 
the bottom shape is preserved as an integral of motion. 

Numerical experiments have been carried out that confirm advantage of the
theory by giving quite realistic pictures for wave profiles before wave 
breaking.

Of course, the above ``inviscid theory'' works only on large enough spatial 
scales and only until the singularity moment, as it was
clear from the very beginning. Practically, this theory is good for description 
sea and ocean waves before their breaking.

\subsection*{Acknowledgments}
These investigations were supported by the INTAS under grant number 00292, 
by RFBR,  by the Russian State Program for Support of the Leading 
Scientific Schools, and by the Russian Science Support Foundation.

\end{document}